\documentclass[pra, twocolumn,showpacs,nofootinbibfloatfix,amsmath,amsfonts,amssymb]{revtex4-1}%
\usepackage{amsmath,amsfonts,amssymb,color}
\usepackage{amsthm}
\usepackage{leftidx}
\usepackage{graphicx}
\usepackage{xcolor}
\usepackage{dcolumn}
\usepackage{bm}
\usepackage{epstopdf}
\usepackage{epsfig}
\usepackage{mathdots} 
\def\RB{\textcolor{black}}
\def\RBn{\textcolor{black}}

\begin{document}
	\title{Quantum-vacuum-\RB{protected} topological edge polaritons}
	\author{Raditya Weda Bomantara}
	\email{Raditya.Bomantara@kfupm.edu.sa}
	\affiliation{%
		Department of Physics, Interdisciplinary Research Center for Intelligent Secure Systems, King Fahd University of Petroleum and Minerals, 31261 Dhahran, Saudi Arabia
	}
	\date{\today}
	
	%%%%%%%%%%%%%%%%%%%% ABSTRACT %%%%%%%%%%%%%%%%%%%%%%%%
	%\begin{linenumbers}
	
	\vspace{2cm}
	
\begin{abstract}
This paper uncovers the formation of topological edge polaritons that are \RB{protected} by the presence of quantum vacuum. Such quantum-vacuum-\RB{protected} edge polaritons could be achieved in a system of \RBn{neutral atomic} lattice under appropriate interaction with a single photonic mode. In the absence of the light-matter coupling, the system is shown to be topologically trivial, which consequently does not support edge modes. By employing Floquet theory, the system is also found to be topologically trivial in the classical light limit, i.e., at very small light-matter coupling but very large number of photons. On the other hand, by treating both the \RBn{atomic} and photonic degrees of freedom quantum mechanically, the system becomes topologically nontrivial in the full (\RBn{atomic}+photonic) Hilbert space, which manifests itself as a pair of topological (almost) zero energy eigenstates localized near each lattice's edge and with very small mean photon number. Finally, the robustness of such quantum-vacuum-\RB{protected} edge polaritons against spatial disorder and counterrotating coupling effect is explicitly demonstrated.
	
\end{abstract}

\maketitle

\section{Introduction} 
\label{intro}

Quantum vacuum has remained a central subject in the area of quantum electrodynamics that underlies various physical phenomena such as the Casimir effect \cite{Casimir48}, spontaneous emission \cite{Frisch33}, the Lamb shift \cite{Lamb47}, and (very recently) quantum-vacuum-induced superconductivity \cite{Schlawin19}. In particular, these phenomena arise due to the interaction between particles and virtual photons, which are induced by quantum vacuum fluctuations that result from the presence of such particles. Observing the quantum effects of light, including quantum vacuum, involves access to the regime of strong light-matter coupling in systems of particles that interact with a sufficiently low number of photons. This is typically achieved by placing such particles inside an optical cavity/quantum resonators, which provides a controllable platform for trapping particles and photons, as well as for engineering appropriate interaction among them. Such a physical setup has been extensively studied under the timely area of cavity quantum electrodynamics \cite{Sentef18,Vidal21,Schlawin22,Torre21}, and its implementation in the lab has been successfully demonstrated \cite{Landig16,Chikkaraddy16,Santhosh16}.

This work uncovers another mechanism under which quantum vacuum yields a physical observable with no classical counterpart. Specifically, by noticing that quantum vacuum mimics a physical edge of a one-dimensional (1D) system, it is capable of supporting the analogues of zero energy edge modes (as typically found in 1D topological insulators \cite{Asboth16} and topological superconductors \cite{Kit}), under an appropriately constructed Hamiltonian. It is worth noting that zero energy edge modes are particularly attractive for quantum technological applications. Indeed, a pair of the so-called Majorana zero modes that emerge at the ends of 1D topological superconductors collectively form a nontrivial qubit, which consequently exhibits natural protection against physical errors \cite{Nayak08}. For this reason, efforts towards realizing Majorana-based qubits in experiments \cite{Lutchyn10,Oreg10,Mourik12,Nichele17,Das12,Rokhinson12,Lee14} and harnessing their quantum computational power \cite{Karzig17,Litinski17,Landau16,Plugge16,Litinski18,Viyuela19} have been extensively made since over the last two decades and are still ongoing \cite{Paetznick23}. Non-Majorana zero modes, such as those that emerge at the ends of 1D topological insulators or their bosonic variations, also find applications in other aspects of quantum technologies despite their lack in nonlocal properties. For example, such modes were recently utilized as additional degrees of freedom to encode quantum information locally at one end of a topological quantum circuit \cite{Mei2018}. By adiabatically tuning appropriate system parameters, the encoded quantum information could then be transmitted to the other end of the circuit with high fidelity \cite{Mei2018}, thus achieving the process of quantum state transfer. 

The ability of light to induce topological features in an otherwise trivial system may be reminiscent of Floquet topological phases \cite{Lindner11,Jiang11}. However, it is worth noting that the mechanism underlying our proposed system in this paper is fundamentally different from that underlying Floquet topological phases. In particular, Floquet topological phases arise when light interacts {\it classically} with the system at hand \cite{Lindner11,Jiang11}, thus rendering its Hamiltonian time-periodic. As a result of this time-periodicity, energy is no longer conserved and is replaced by the so-called quasienergy, which is only defined modulo the driving frequency \cite{Shirley65,Sambe73}. Due to the periodicity of the quasienergy space, additional band touching events may arise and consequently induce a topological phase transition, thus resulting in a variety of novel topological phases, often with no static counterparts \cite{Bomantara16,Rudner13,Nathan15,Asboth14,Ho14,Bomantara20,Zhou18,Zhu22,Bomantara22,Koor22}. Apart from the emergence of band touching events at the edges of the quasienergy Brillouin Zone, Floquet topological phases may also arise when the (classical) light-matter interaction renormalizes some system parameters in the effective static Hamiltonian description, thus rendering the system topologically nontrivial even when its undriven counterpart is topologically trivial \cite{Ino10,Kitagawa10,Oka09,Calvo11,Kitagawa11,Morell12}. By contrast, the topological effect arising in the present work is not caused by any of the aforementioned mechanisms, but instead results from the fact that the photonic Fock space is bounded from below by the quantum vacuum. Indeed, without taking into account the quantum effect of light, such a topological feature will not be present.

Finally, it is worth noting that studies of Floquet topological phases under full quantum light treatment have developed into an active research area since the last decade \cite{Trif12,Gulacsi15,Bomantara16b,Ciuti21,Chio21,Dmytruk22,Mendez20,Winter23,Bacciconi24,Karzig15,Wang19}. However, to our knowledge, the role of quantum vacuum in manifesting topological effects has remained elusive. In particular, while the topology of quantum vacuum itself has been studied in the context of high energy physics \cite{Volovik13,Alexander20}, its implication on a physical observable has yet to be identified. This work aims to close this knowledge gap by explicitly demonstrating the presence of topological edge modes in a system with strong light-matter coupling at very low photon mean numbers. This is achieved by devising an appropriate 1D topologically trivial lattice \RBn{of neutral atoms or artificial atoms} that interacts with a single photonic mode via a Tavis-Cummings-like coupling, \RBn{which could arise in cold atom or superconducting circuit systems}. Such an interaction renders the composite light-matter system (polariton) topologically nontrivial in the enlarged Hilbert space containing the \RBn{atomic} and the photonic degrees of freedom. Remarkably, as the quantum vacuum state behaves as an ``edge" in the photonic Fock space, the system topology manifests itself as in-gap states that are only present near the lattice's (physical) edges {\it and} at very low photon mean numbers. That is, such edge states are not observable in the classical light limit and their existence is linked to the presence of quantum vacuum, as intended. \RB{It should however be emphasized that we take a broad definition of quantum vacuum in this paper, which does not necessarily correspond to a true electromagnetic vacuum. The notion of quantum vacuum used in this paper may generally refer to the zero number state that serves as the boundary of any effective Fock space.}

This manuscript is organized as follows. In Sec.~\ref{model}, we mathematically present the Hamiltonian describing the proposed system. In Sec.~\ref{intuition}, we elucidate the intuition behind the system's construction by explicitly demonstrating the emergence of quantum-vacuum-\RB{protected} topological edge polaritons at special parameter values. In Sec.~\ref{sec:general}, we numerically calculate the system's energy spectrum and identify the quantum-vacuum-\RB{protected} topological edge polaritons at more general parameter values. There, we further verify that the quantum-vacuum-\RB{protected} topological edge polaritons have an almost zero mean photon number. In Sec.~\ref{class}, we verify the absence of any topological edge mode in the system when the light-matter coupling is treated classically. In Sec.~\ref{discuss}, we demonstrate the robustness of our quantum-vacuum-\RB{protected} topological edge polaritons against spatial disorder and counterrotating coupling term in the system's Hamiltonian. Finally, we summarize our results and discuss prospects for future studies in Sec.~\ref{conc}.

\section{Topological edge modes protected by quantum vacuum}

\subsection{Model description}
\label{model}

We consider a 1D chain of \RBn{two-level neutral atoms or artifical atoms} interacting with a photon mode under the Hamiltonian ($\hbar =1$ units are used throughout this paper)
	\begin{equation}
	H=H_{\rm atom}+ H_{\rm ph} + H_{\rm int} \;, \label{SSH_vac}
	\end{equation}
where 
\begin{eqnarray}
    H_{\rm atom} &=& \sum_{s=\pm 1} \left\lbrace\sum_{j=1}^N J_1 c_{s,A,j}^\dagger c_{s,B,j} + \sum_{j=1}^{N-1} J_2 c_{s,B,j}^\dagger c_{s,A,j+1} \right. \nonumber \\
    && \left. +h.c.  \right\rbrace + \sum_{j=1}^N \sum_{S=A,B} E_Z \left(c_{+,S,j}^\dagger c_{-,S,j} +h.c.  \right) \;, \nonumber \\
    H_{\rm ph} &=& \omega a^\dagger a \;, \nonumber \\
    H_{\rm int} &=& \sum_{j=1}^N \Delta \left\lbrace a \left( c_{+ , A,j}^\dagger c_{-, B , j}  + c_{+ , B,j}^\dagger c_{-, A , j}\right) +h.c. \right\rbrace  \;, \nonumber \\
\end{eqnarray}
$c_{s,X,j}$ ($c_{s,X,j}^\dagger$) is the particle annihilation (creation) operator with its two internal levels labeled as $s=\pm 1$, sublattice $X\in\left\lbrace A,B\right\rbrace$, and at the $j$th site of $N$ lattice sites, $a$ ($a^\dagger$) is the bosonic annihilation (creation) operator, $\omega$ is the corresponding photon energy, $J_1$ and $J_2$ are respectively the intra- and inter-site hopping amplitudes, $E_Z$ is the (artificial) magnetic field strength, and $\Delta$ is the light-matter interaction strength.

$H_{\rm atom}$ describes a pair of the Su-Schrieffer-Heeger (SSH) models \cite{Su1980}, each of which is known to support a topological phase. Indeed, in the special case of $J_1=E_Z=0$, the edge operators $c_{\pm ,A, 1}$ and $c_{\pm ,B, N}$ commute with $H_{\rm atom}$, consequently representing its zero modes. At $J_1\neq 0$ but with $E_Z=0$, zero modes remain present as long as $J_1<J_2$, though they no longer correspond to $c_{\pm ,A, 1}$ and $c_{\pm ,B, N}$ anymore. It should be noted, however, that such zero modes cease to exist as $E_Z$ becomes nonzero. In particular, the magnetic field term of $H_{\rm atom}$ hybridizes the zero modes localized on the same edge but associated with different internal levels, thereby lifting their degeneracy. Therefore, the Hamiltonian $H_{\rm atom}$ as a whole is {\it not} topologically protected, despite it being made up of two topological models.    

$H_{\rm ph}$ describes the photonic part of the Hamiltonian, whilst $H_{\rm int}$ describes the interaction part that couples the \emph{atomic} and photonic degrees of freedom. The interaction term could be interpreted as a modified Tavis-Cummings (TC) term \cite{Tavis68} which involves an atom hopping from one sublattice to the other in addition to flipping its internal level state. Note that in potential experimental realizations of our model, the internal levels could be represented by the ground and excited states of a two-level superconducting qubit \cite{Fink09}. In this work, we assume the framework of a single \RBn{neutral} particle occupying our 1D lattice, which in turn significantly reduces the complexity of our model. While this eliminates the observation of interesting phenomena typically associated with the TC model such as superradiance \cite{Tavis68}, it suffices to demonstrate the topological effects \RB{protected} by quantum vacuum.  

In the following, we will show that the constructed interaction term above causes a pair of topologically protected zero modes with a very small photon number expectation value to emerge. This is made possible by the presence of quantum vacuum through a mechanism that is reminiscent of how topological modes arise at the edges of a topologically nontrivial system. 

\subsection{Intuitive insight at special parameter values} 
\label{intuition}

We first specialize to the case of $J_1=E_Z=0$. The operator 
\begin{equation}
    \mathcal{S} = \sum_{j=1}^N \sum_{s=\pm 1} \sum_{X=A,B} \frac{s}{2} c_{s,X,j}^\dagger c_{s,X,j} +a^\dagger a 
\end{equation}
then serves as a conserved quantity with respect to the full Hamiltonian $H$ (the eigenvalue $\ell=n+s/2$ of $\mathcal{S}$ is a good quantum number). Consequently, the set of states $\left\lbrace |+,n,X,j\rangle , |-,n+1,X,j\rangle : X\in \left\lbrace A,B \right\rbrace , j=1,\cdots, N\right\rbrace$, where $c_{\pm,X,j}^\dagger c_{\pm,X,j} |\pm,n,X,j\rangle = |\pm,n,X,j\rangle$ and $a^\dagger a |\pm,n,X,j\rangle = n |\pm,n,X,j\rangle$, is closed under the application of $H$.

\begin{center}
\begin{figure}
    \includegraphics[scale=0.3]{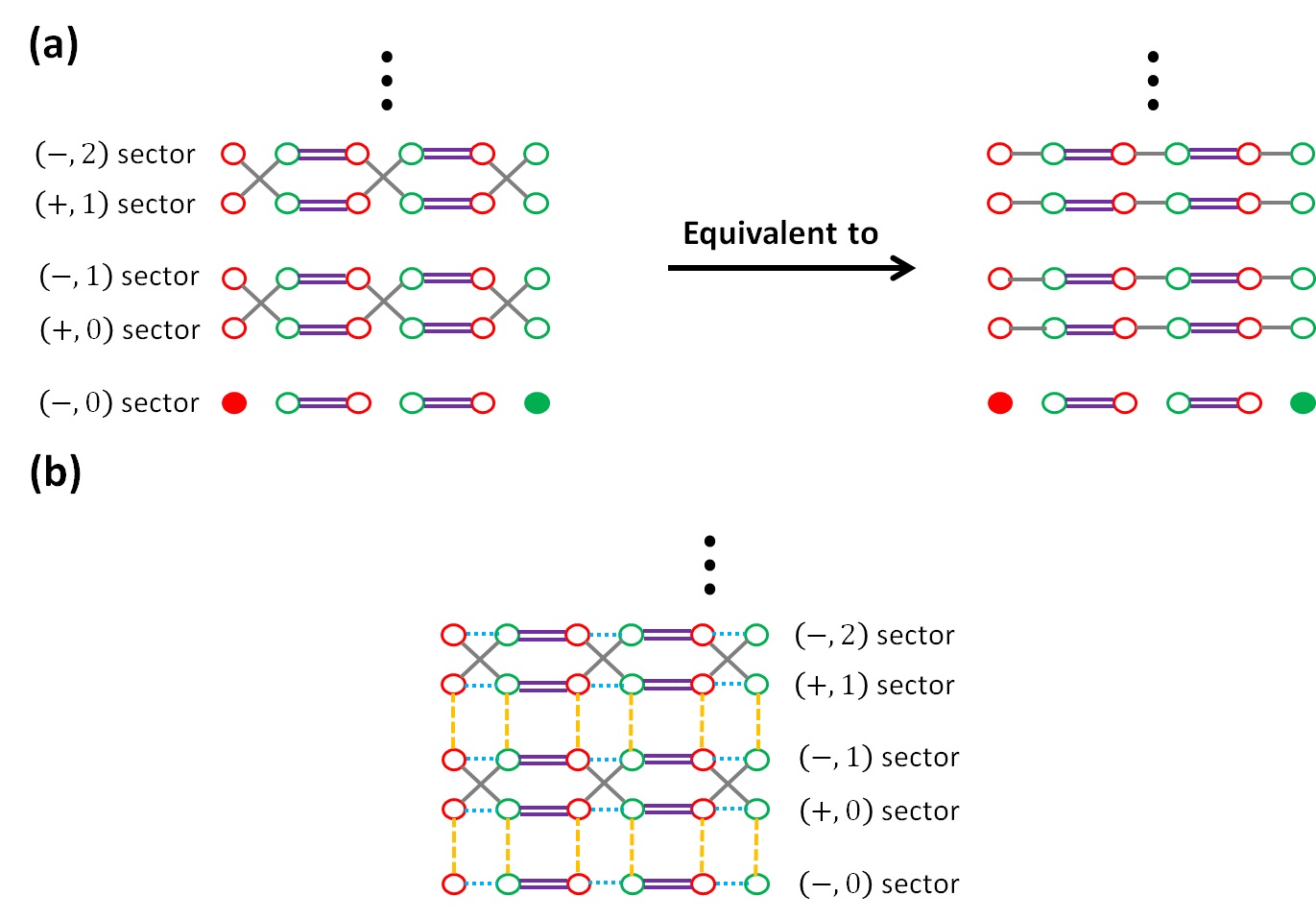}
    \caption{The schematics of our system's Hamiltonian in terms of the single particle states $|s,n,S,j\rangle$ under (a) the special case of $J_1=E_Z=0$ and (b) general non-zero parameter values. The red and green circles describe the sublattice A and B of the system respectively. The states corresponding to different $(s,n)$ sectors are aligned in the $y$-direction. In panel (a), the filled circles denote the zero modes.}
    \label{fig:scheme}
\end{figure}    
\end{center}

Figure~\ref{fig:scheme}(a) shows the schematics of our Hamiltonian under the special case $J_1=E_Z=0$ presented above, from which two important observations are made. First, the $(-,n+1)$ and $(+,n)$ atom-photon sectors coupled by the Hamiltonian can be deformed into two independent SSH models with an effective intra-(inter-)site coupling strength of $\Delta \sqrt{n+1}$ ($J_2$) and with an effective sublattice dependent onsite potential that is $\propto \omega$. Consequently, a pair of edge modes might exist at each lattice edge for atom-photon sectors that satisfy $\Delta \sqrt{n+1} <J_2$. Nevertheless, such edge modes are not topological even at these special parameter values due to the presence of the onsite potential. Moreover, for $\Delta \gg J_2$, such edge modes are absent for any pair of $(-,n+1)$ and $(+,n)$ atom-photon sectors. Second, note that the $(-,0)$ atom-photon sector is detached from the other sectors and behaves as an SSH model in the topologically nontrivial regime regardless of the light-matter interaction strength. As a result, at the special parameter values under consideration, the full Hamiltonian $H$ \textit{always} supports a pair of zero modes in the $(-,0)$ atom-photon sector, which correspond to the states $|- , 0, A,1 \rangle$ and $|-, 0, B,N \rangle$. 

\RB{Figure~\ref{fig:scheme}(b) shows the schematics of our Hamiltonian when all system parameters are nonzero. In this case, the system is no longer separated into decoupled sectors and instead forms an effective 2D lattice spanned by the combined atomic and photonic degrees of freedom, in which \emph{all} sites are nontrivially connected. This effective 2D lattice has a qualitatively similar structure to a typical second-order topological insulator \cite{STI1,STI2,Bomantara19,STI3,STI4,STI5,STI6,STI7}, which is known to support topologically protected corner modes. However, due to the the dependence of the coupling in the ``photon number direction" on $\sqrt{n+1}$, the effective 2D Hamiltonian of our system is not exactly the same as that of any existing second-order topological insulator, and its full topological characterization is much more difficult. Nevertheless, upon identifying the presence of corner modes under the special parameter values above, the system's second-order topology could still be established by verifying that such corner modes are indeed robust at more general parameter values as detailed in the following. To gain further intuition along the way, we will start by analyzing the simplest generalization by setting $E_Z=0$ and $J_1\neq 0$, under which the system is still separated into decoupled sectors. We will then consider the slightly more complicated case of $E_Z\neq 0$ and $J_1=0$, under which all sites are coupled and the full effective 2D lattice must be accounted for. Finally, the most general case of all system parameters being nonzero will be discussed in Sec.~\ref{sec:general}.}

\vspace{0.5cm}
\noindent {\it Case i:} $E_Z= 0$ and $J_1\neq 0$

\vspace{0.5cm}
In this case, $\ell=n+s/2$ remains a good quantum number with which the Hamiltonian can be divided into decoupled sectors via
%\begin{eqnarray}
%    \mathcal{H}_{\rm eff, k} &=& (J_1 +J_2 \cos(k)) \sigma_x +J_2 \sin(k) \sigma_y +\sqrt{n+1} \Delta \tau_x \sigma_x \nonumber \\
%    && + \frac{\omega}{2} \tau_z , 
%\end{eqnarray}
\begin{eqnarray}
    H &=& \sum_{\ell=-1/2,0,1/2,\cdots }^\infty  \mathcal{H}_{\rm eff, \ell} |\ell \rangle \langle \ell |  ,  
\end{eqnarray}
where $\langle \ell | = \left( \langle -,0,X,j | \right)$ for $\ell=-1/2$ and $\langle \ell | = \left( \langle +,n,X,j | , \langle -,n+1,X,j | \right)$ for $\ell = n+1/2$, $X=A,B$, and $j=1,2,\cdots , N$. The effective $2\times 2$ Hamiltonian describing each decoupled sector $\ell$ can be written as
\begin{eqnarray}
    \mathcal{H}_{\rm eff, \ell = -1/2} &=& \left[J_1 +J_2 \cos(\hat{k}) \right] \sigma_x + J_2 \sin(\hat{k}) \sigma_y , \nonumber \\
    \mathcal{H}_{\rm eff, \ell\neq -1/2} &=& \left[ J_1 +J_2 \cos(\hat{k}) \right] \sigma_x +J_2 \sin(\hat{k}) \sigma_y  \nonumber \\
    && +\sqrt{\ell+1/2} \Delta \tau_x \sigma_x + \frac{\omega}{2} \tau_z , \label{heffbulk}
\end{eqnarray}
where we have suppressed unimportant constant terms, $\hat{k}$ is the quasimomentum operator such that $e^{\pm \mathrm{i} \hat{k}}$ shifts the lattice site ($j$) by one unit, $\sigma$'s and $\tau$'s are respectively the Pauli matrices acting on the sublattice ($A,B$) and sector-$\ell$ ($(+,n),(-,n+1)$) degrees of freedom. 
 
 $\mathcal{H}_{\rm eff, \ell = -1/2}$, which corresponds to the $(-,0)$ atom-photon sector, remains mathematically identical to the paradigmatic SSH model and thus continues to support a pair of zero modes as long as $J_1<J_2$. The remaining pairs of $(-,n+1)$ and $(+,n)$ atom-photon sectors are effectively described by $\mathcal{H}_{\rm eff, \ell \neq -1/2}$, which also satisfies the chiral symmetry $\mathcal{C}^\dagger \mathcal{H}_{\rm eff, \ell}(\hat{k}) \mathcal{C} = -\mathcal{H}_{\rm eff, \ell}(\hat{k})$, particle-hole symmetry $\mathcal{P}^\dagger \mathcal{H}_{\rm eff, \ell}(\hat{k}) \mathcal{P} = -\mathcal{H}_{\rm eff, \ell}(-\hat{k})$, and time-reversal symmetry $\mathcal{T}^\dagger \mathcal{H}_{\rm eff, \ell}(\hat{k}) \mathcal{T} = \mathcal{H}_{\rm eff, \ell}(-\hat{k})$, but under different operators $\mathcal{C} = \tau_x \sigma_z$, $\mathcal{P} = \mathcal{K} \tau_x \sigma_z$, and $\mathcal{T} = \mathcal{K}$ respectively, where $\mathcal{K}$ is the complex conjugation operator. Therefore, all $\mathcal{H}_{\rm eff, \ell}$ belongs to the BDI class according to the Altland-Zirnbauer classification \cite{AZ97}, the topology of which is characterized by a $\mathbb{Z}$ invariant.  
 
 %Unlike $\mathcal{H}_{\rm eff, \ell = -1/2}$, however, $\mathcal{H}_{\rm eff, \ell \neq -1/2}$ does not exhibit a topological phase transition at $J_1=J_2$. Indeed, as demonstrated in Fig.~\ref{fig:windbulkpresults}(a), no gap closing is observed in the energy spectrum of $\mathcal{H}_{\rm eff, \ell \neq -1/2}$ at $J_1=J_2$. Consequently, it remains topologically trivial and does not support edge modes at all values of $J_1$.   

\begin{center}
\begin{figure}
    \includegraphics[scale=0.22]{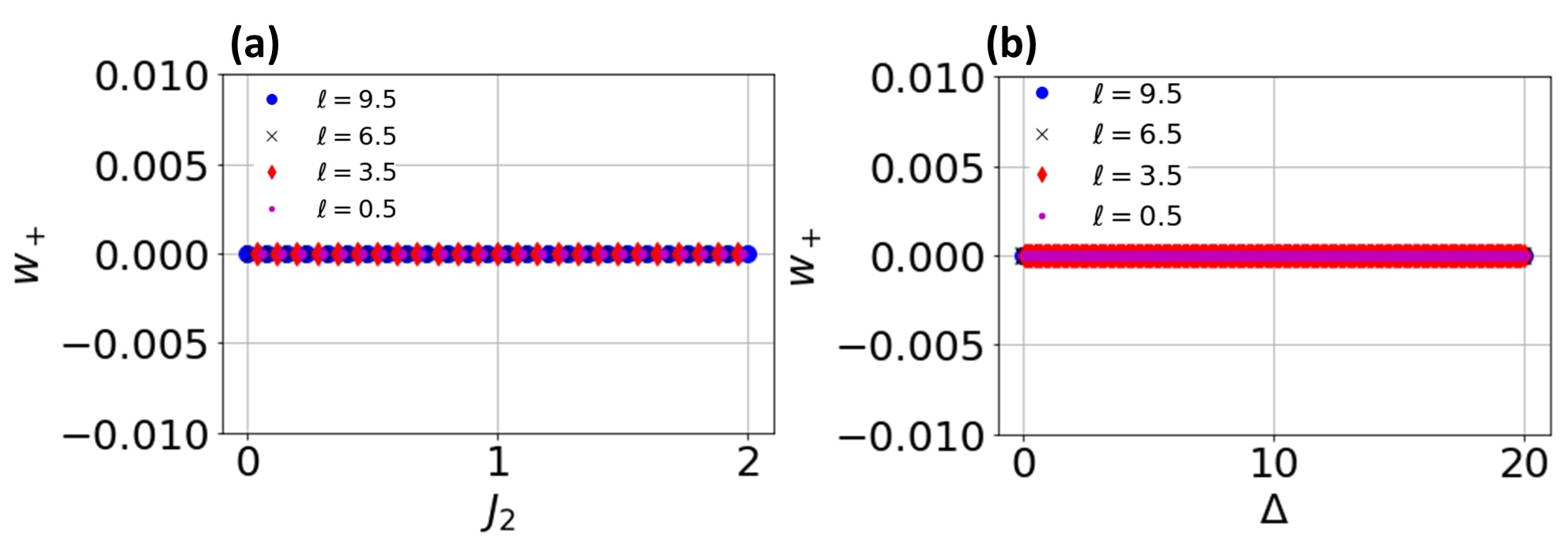}
    \caption{\RB{The winding number $w_+$ (a) as a function of $J_2$ at a fixed $\Delta=J_1=1$ and $\omega =2\pi$, and (b) as a function of $\Delta$ at a fixed $J_2=0.5$, $J_1=0.1$, and $\omega =2\pi$.}}
    %\caption{(a) The energy spectrum of Eq.~(\ref{heffbulk}) at varying $J_1$ for different values of $\ell$ (energy shift of $\ell \omega$ has been reintroduced to account for the suppressed constant terms). (b) The winding number $w_+$ as a function of the light-matter coupling $\Delta$. In both panels, the parameter values are set at $J_2=0.5$, $\omega =2\pi$, (a) $\Delta =10$, and (b) $J_1=0.1$}
    \label{fig:windbulkpresults}
\end{figure}    
\end{center}

In the canonical basis where $\mathcal{C} \rightarrow \left( \begin{array}{cc}
 \mathbf{I}   &   \mathbf{0}  \\
 \mathbf{0}    &    -\mathbf{I}
\end{array}\right) $ is block diagonal in the sublattice basis, we have
\begin{equation}
    \mathcal{H}_{\rm eff, \ell \neq -1/2}(\hat{k}) \hat{=} \left(\begin{array}{cc}
      \mathbf{0}   & h_-   \\
      h_+   & \mathbf{0}
    \end{array}\right) ,
\end{equation}
where
\begin{equation}
    h_\pm = J_1+J_2 \cos(\hat{k}) \pm \mathrm{i} J_2 \sin(\hat{k}) \tau_x +\sqrt{\ell+1/2} \Delta \tau_x +\frac{\omega}{2} \tau_z . 
\end{equation}
We could then define the topological invariant for $\mathcal{H}_{\rm eff, \ell \neq -1/2}$ to be the winding number 
\begin{equation}
    w_\pm=\frac{1}{2\pi \mathrm{i}} \oint \mathrm{Tr} \left(h_\pm^{-1} dh_\pm  \right) ,
\end{equation}
which is defined under the assumption of periodic boundary conditions (PBC) such that $\hat{k}\rightarrow k \in [-\pi,\pi)$ is a good quantum number. We numerically compute the winding number $w_+$ at four different $\ell$ sectors in Fig.~\ref{fig:windbulkpresults}. \RB{In particular, by varying $J_2$ while fixing $J_1=\Delta=1$, Fig.~\ref{fig:windbulkpresults}(a) reveals that $w_+$ is trivial everywhere, indicating the absence of a topological phase transition at $J_1=J_2$ in all these nonzero photon sectors.} Figure~\ref{fig:windbulkpresults}(b) further demonstrates that $w_+$ is also $0$ for all $\Delta$ values under consideration, even in the regime $J_1<J_2$ where each SSH model is topologically nontrivial. While not shown in the figure, the same results are also observed for $w_-$. Our winding number calculations thus confirm the absence of topological edge modes at $\ell\neq -1/2$. Therefore, under $E_Z=0$, the system only supports a pair of topological edge modes in the quantum vacuum sector of the Hamiltonian, i.e., as described by $\mathcal{H}_{\rm eff, \ell = -1/2}$.  

\vspace{0.5cm}
\noindent {\it Case ii:} $E_Z\neq 0$ and $J_1=0$

\vspace{0.5cm}
In this case, the $(-,0)$ atom-photon sector is no longer decoupled. However, by additionally setting $\omega=0$ for simplicity, we find that a pair of zero modes still exist, but now correspond to the superposition states which can be found up to first order in $J_2$ as (see Appendix~\ref{app:A} for details)
\begin{eqnarray}
   |0_L\rangle &=& \sum_{n=0}^\infty \left[f(2n+1) |-,2n+1,B,1\rangle  \right. \nonumber \\
   && \left. - f(2n) |-,2n,A,1\rangle \right] +\mathcal{O}(J_2), \nonumber \\
   |0_R\rangle &= & \sum_{n=0}^\infty \left[f(2n) |-,2n,B,N\rangle  \right. \nonumber \\
   && \left.- f(2n+1) |-,2n+1,A,N\rangle \right] +\mathcal{O}(J_2) , 
\end{eqnarray}
where 
\begin{equation}
    f(n) = \left(\frac{E_Z}{\Delta}\right)^n \frac{1}{\sqrt{n!}} .
\end{equation}
It is worth noting that a function of the form $x^n/\sqrt{n!}$ converges to $0$ as $n\rightarrow \infty$ regardless of $x$. As such, $|0_L\rangle$ and $|0_R\rangle$ are normalizable for any finite value of $E_Z$ (provided $\Delta \neq 0$). This demonstrates the robustness of the system's edge modes against the presence of the perturbation $E_Z$ that connects all atom-photon sectors.

\subsection{The emergence of topological edge modes at general parameter values}
\label{sec:general}

\begin{center}
\begin{figure*}
    \includegraphics[scale=0.5]{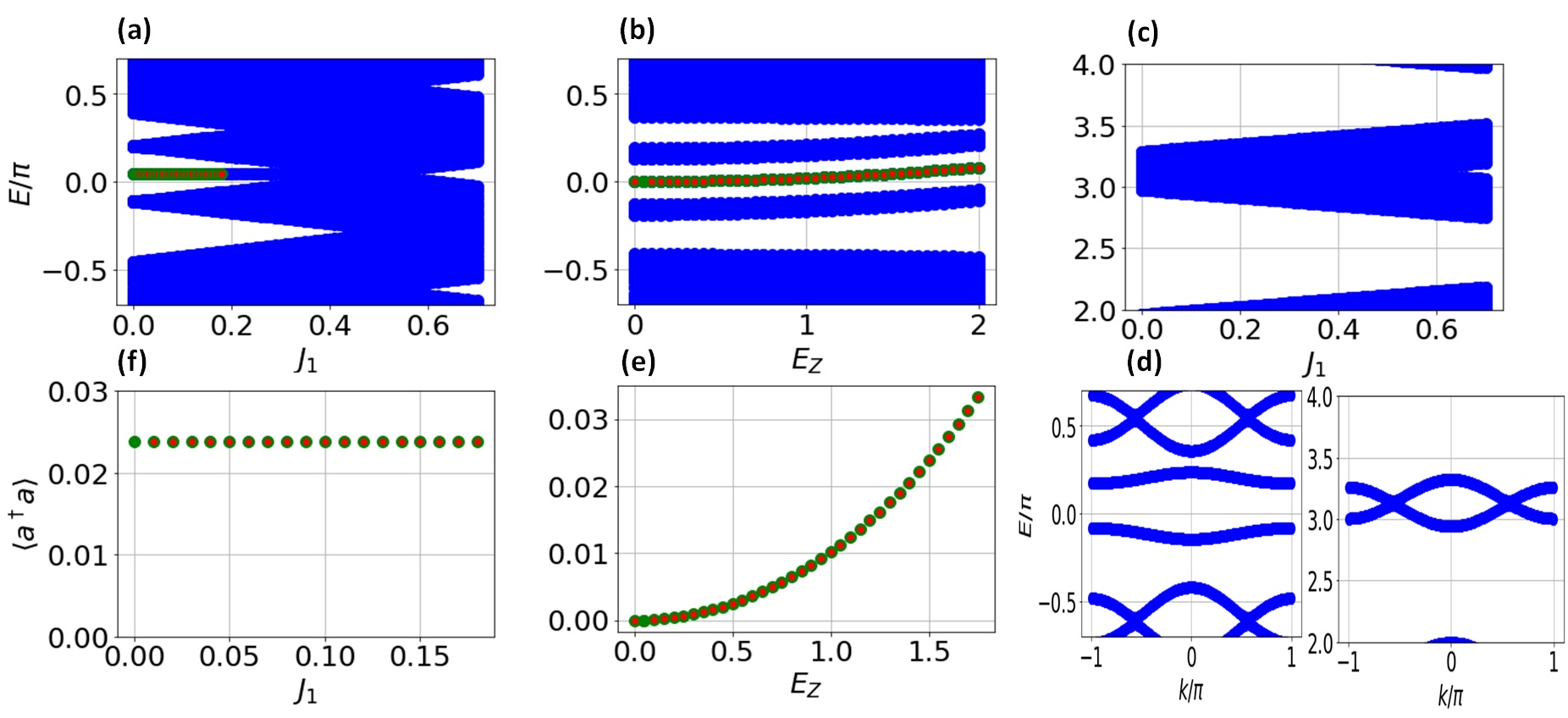}
    \caption{(a,b,c) The energy spectrum of Eq.~(\ref{SSH_vac}) for $N=20$ at (a,c) varying $J_1$ and fixed $E_Z=1.5$, (b) varying $E_Z$ and fixed $J_1=0.1$. (d) The corresponding momentum space energy band structure of the system in a closed geometry at a fixed $J_1= 0.1$ and $E_Z=1.5$, plotted around two energy references ($E=0$ and $E=3$) for clarity. (e,f) The photon number expectation values of the edge states at varying (e) $E_Z$ and (f) $J_1$. The left-edge-(right-edge-)localized eigenstates are marked in green (red) respectively. In all panels, the other parameters are set as $J_2=0.5$, $\Delta =10$, $\omega =2\pi$, and the Fock space is truncated to contain up to $N_p=20$ photon sectors.}
    \label{fig:realspresults}
\end{figure*}    
\end{center}

In Figs.~\ref{fig:realspresults}(a) and (b), we show the few energy levels closest to $\langle a^\dagger a \rangle =0$ at varying intra-site hopping and magnetic field strength respectively. There, \RB{all system parameters are nonzero, which means that all atom-photon sectors are nontrivially coupled. Nevertheless,} left- and right-edge-localized eigenstates could be clearly identified inside the gap between the two bulk energy levels closest to $E=0$. As demonstrated in Fig.~\ref{fig:realspresults}(c), such edge states are absent at higher energy levels that correspond to larger $\langle a^\dagger a \rangle $. This serves as strong evidence that the obtained edge states are indeed \RB{protected} by the quantum vacuum. \RB{Moroever, it should be emphasized that while such edge states are generally no longer found at zero energy, they remain degenerate and are adiabatically connected to the zero energy edge states in the special case of $E_Z=0$, which are therefore topological in nature.}

By considering a closed lattice geometry, Eq.~(\ref{SSH_vac}) can be written in the form
\begin{equation}
    H=\sum_k \psi_k^\dagger \mathcal{H}_k \psi_k,
\end{equation}
where $\psi_k = (c_{-,A,k}, c_{+,A,k}, c_{-,B,k}, c_{+,B,k})^T$, $c_{\pm,S,k} = \frac{1}{\sqrt{N}}\sum_j c_{\pm , S, j} e^{i j k}$ is the particle annihilation operator in the momentum space,
\begin{eqnarray}
    \mathcal{H}_k &=& (J_1+J_2 \cos(k)) \sigma_x + J_2 \sin(k) \sigma_y +E_Z \zeta_x +\omega a^\dagger a \nonumber \\
    && +\Delta (a \zeta_+ + a^\dagger \zeta_-)\sigma_x . \label{momspaceH}
\end{eqnarray}
is the corresponding momentum space Hamiltonian, $\sigma$'s and $\zeta$'s are respectively the Pauli matrices acting on the sublattice and internal degrees of freedom, and $\zeta_\pm =\zeta_x \pm \mathrm{i} \zeta_y$. Upon diagonalizing $\mathcal{H}_k$ as a function of $k$, we found that the two energy bands closest to $E=0$ are gapped (see the left panel of Fig.~\ref{fig:realspresults}(d)), whilst the other pairs of energy bands are gapless (see both panels of Fig.~\ref{fig:realspresults}(d) for some representative gapless bands). This further confirms that topological edge states, if exist, could only emerge in the gap between the two bands closest to the quantum vacuum sector.

Finally, in Figs.~\ref{fig:realspresults}(e) and (f), we plot the photon number expectation value $\langle a^\dagger a \rangle$ associated with the edge states observed in panels (b) and (a) respectively. In both cases, we found that $\langle a^\dagger a \rangle \approx 0$ for all parameter values under consideration, as expected from the fact that such edge states exist near the quantum vacuum sector. That $\langle a^\dagger a \rangle \approx 0$ increases with $E_Z$ in Fig.~\ref{fig:realspresults}(e) is attributed to the fact that the magnetic field term couples the $(-,0)$ atom-photon sector with the other higher photon sectors. Nevertheless, as $\langle a^\dagger a \rangle $ remains close to zero even at moderate values of $E_Z$, such edge states remain localized near the $a^\dagger a =0$ photon sector, thus confirming their topological nature in the same spirit as the robust localization of ordinary topological edge states near a physical edge against considerable perturbations. \RB{Due to the crucial role of quantum vacuum as an artificial edge that facilitates the above obtained edge states, the latter shall be referred to as \emph{quantum-vacuum-\RB{protected} edge polaritons} in the remainder of this paper.}

\section{Absence of topological edge modes in the classical limit}
\label{class}

In the previous section, we have shown that quantum vacuum plays a crucial role in the formation of topological edge states. To further support this result, we will now show that such topological edge states will not be present in the absence of quantum vacuum, i.e., when the photon mode is treated classically. Such a classical limit could be obtained by replacing $a\rightarrow e^{-\mathrm{i} \omega t}$, as well as setting $\Delta \rightarrow 0$ and $n\rightarrow \infty$ while keeping $\Delta \sqrt{n} \rightarrow $ constant in Eq.~(\ref{SSH_vac}). The result is a time-periodic Hamiltonian of the form
\begin{eqnarray}
    H(t) &=& \sum_{s=\pm 1} \left\lbrace\sum_{j=1}^N J_1 c_{s,A,j}^\dagger c_{s,B,j} + \sum_{j=1}^{N-1} J_2 c_{s,B,j}^\dagger c_{s,A,j+1} \right. \nonumber \\
    && \left. +h.c.  \right\rbrace + \sum_{j=1}^N \sum_{S=A,B} E_Z \left(c_{+,S,j}^\dagger c_{-,S,j} +h.c.  \right) \nonumber \\
    && +\sum_{j=1}^N \Delta \left\lbrace e^{-\mathrm{i} \omega t} \left( c_{+ , A,j}^\dagger c_{-, B , j}  + c_{+ , B,j}^\dagger c_{-, A , j}\right) \right. \nonumber \\
    && \left. +h.c. \right\rbrace  \;,
    \label{SSH_class}
\end{eqnarray}
where we have redefined $\Delta \sqrt{n}$ as $\Delta$ in the classical limit.

\begin{center}
\begin{figure}
    \includegraphics[scale=0.22]{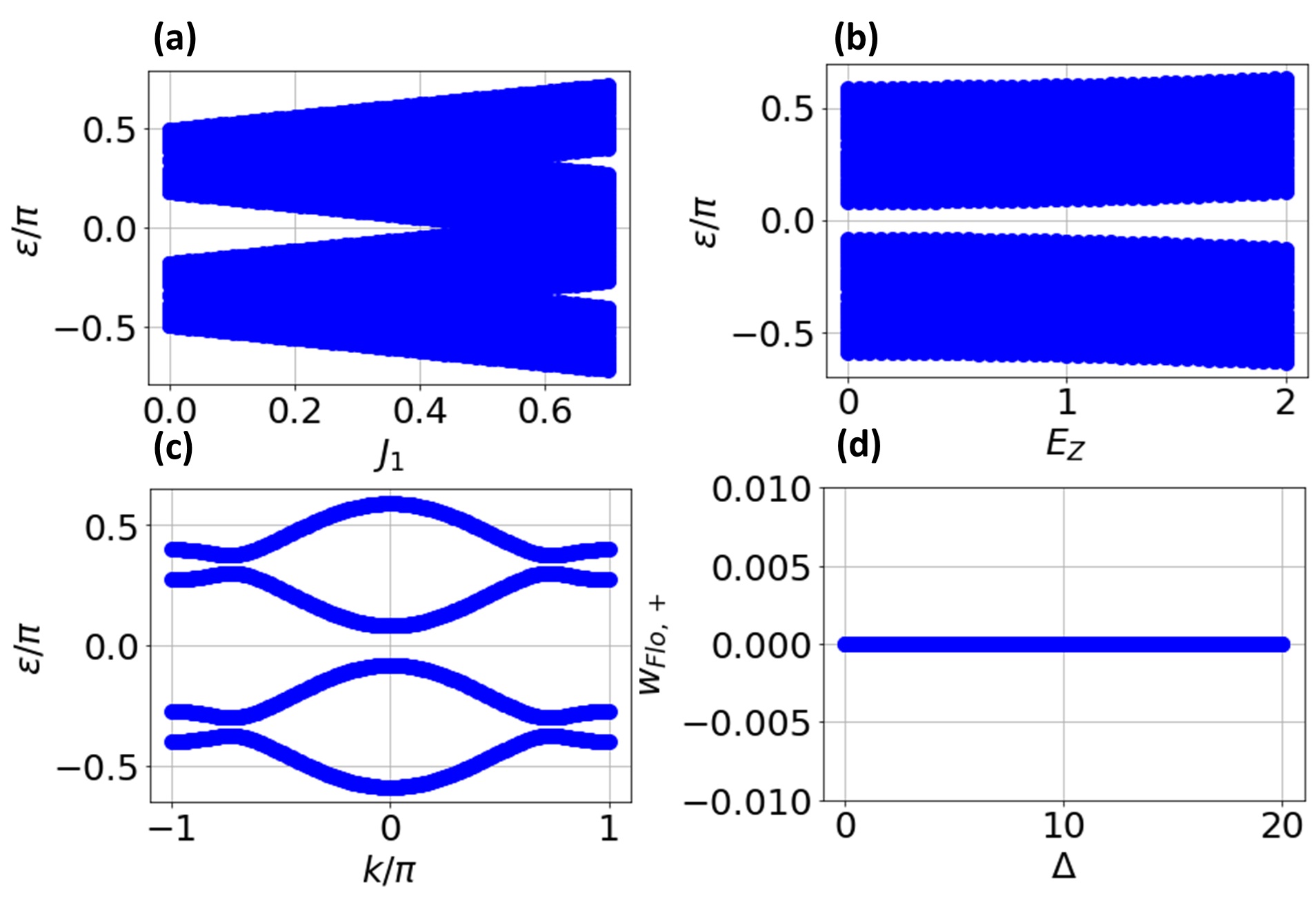}
    \caption{(a,b,c) The quasienergy spectrum associated with the time-periodic Hamiltonian $H(t)$ of Eq.~(\ref{SSH_class}) under (a,b) OBC, and (c) PBC. (d) The numerically computed normalized winding number as a function of the light-matter coupling $\Delta$. The system parameters are chosen as $J_2=0.5$, $\Delta=10$, (a,c,d) $E_Z=0$, (b,c,d) $J_1=0.3$.}
    \label{fig:classresults}
\end{figure}    
\end{center}

To analyze such a time-periodic system, we employ the Floquet theory \cite{Shirley65,Sambe73} in which the notions of energy and its corresponding eigenstate are respectively replaced by quasienergy and its corresponding Floquet eigenstate. Specifically, the quasienergy $\varepsilon$ of the system is obtained from the eigenvalue $e^{-\mathrm{i} \varepsilon \frac{2\pi}{\omega}}$ of the one-period time-evolution operator (also referred to as the Floquet operator) generated by the Hamiltonian. Figure~\ref{fig:classresults}(a,b) shows the typical quasienergy spectrum of Eq.~(\ref{SSH_class}) under open boundary conditions (OBC), which indeed demonstrates the absence of edge states as they would have been marked in green/red color should they exist.

To further support the above result, we will now construct a relevant topological invariant for our time-periodic system and show that it only takes a trivial value. To this end, we assume PBC and write our time-periodic Hamiltonian as 
\begin{equation}
    H(t) =\sum_k \psi_k^\dagger \mathcal{H}_k(t) \psi_k,
\end{equation}
where $\psi_k $ is as defined in Sec.~\ref{sec:general} and the time-periodic momentum space Hamiltonian is the same as Eq.~(\ref{momspaceH}) but with the replacement $a\rightarrow e^{-i\omega t}$. By writing the time-dependent Schr\"{o}dinger equation in the frequency space with respect to a quasienergy $\varepsilon$ eigenstate \cite{Shirley65,Sambe73}, we obtain the eigenvalue equation,
\begin{equation}
    \varepsilon |\tilde{\varepsilon}\rangle = \mathcal{H}_{k,Flo} |\tilde{\varepsilon}\rangle ,  
\end{equation}
where
\begin{eqnarray}
    \mathcal{H}_{k,Flo} &=& \frac{\omega}{2} \left( \xi_0+\xi_z \right)+(J_1 +J_2 \cos(k)) \sigma_x +J_2 \sin(k) \sigma_y   \nonumber \\
    && +E_Z \zeta_x + 2\Delta \zeta_x \xi_x \sigma_x + 2\Delta \zeta_y \xi_y \sigma_x
\end{eqnarray}
is the Floquet Hamiltonian associated with $\mathcal{H}_k$ and $\xi$'s are a set of infinite dimensional matrices acting on the frequency space, whose exact expressions are \cite{Bomantara20,Bomantara23}
\begin{eqnarray}
    \left[\xi_0\right]_{a,b} = \delta_{a,b} &,& [\xi_x]_{a,b} = \delta_{a,b+1} + \delta_{a,b-1} , \nonumber \\
    \left[\xi_z\right]_{a,b} = (2b-1)\delta_{a,b} &,& [\xi_y]_{a,b} = \mathrm{i} (\delta_{a,b+1} - \delta_{a,b-1}) .
\end{eqnarray}
The quasienergy bands of $\mathcal{H}_{k,Flo}$ within the first Floquet Brillouin zone, i.e., $\varepsilon \in \left(-\omega/2 , \omega/2 \right]$ are numerically computed in Fig.~\ref{fig:classresults}(c) for completeness.

It is useful to define another set of infinite dimensional matrices 
\begin{eqnarray}
    \left[\eta_x\right]_{a,b} =\delta_{1-a,b}\;, &\left[\eta_y\right]_{a,b} =(-\mathrm{i})^{2b-1} \delta_{1-a,b}\;,& \left[\eta_z\right]_{a,b} =(-1)^b \delta_{a,b}  \nonumber \\
\end{eqnarray}
which are unitary and satisfy $\eta_j \xi_i \eta_j =(2\delta_{i,j}-1)\xi_i$ \cite{Bomantara20,Bomantara23}. By ignoring the constant term, $
\mathcal{H}_{k,Flo}$ satisfies all chiral, particle-hole, and time-reversal symmetries under the operators $\mathcal{C}=\eta_y \zeta_y \sigma_z$, $\mathcal{P}=\mathcal{K} \eta_y \zeta_y \sigma_z$, $\mathcal{T}=\mathcal{K}$ respectively, thus again placing the system in the BDI class that is characterized by a $\mathbb{Z}$ invariant. In the canonical basis where $\mathcal{C}\rightarrow  \left( \begin{array}{cc}
 \mathbf{I}   &   \mathbf{0}  \\
 \mathbf{0}    &    -\mathbf{I}
\end{array}\right) $ is block-diagonal in the sublattice basis, the momentum space Floquet Hamiltonian can be written in the form
\begin{equation}
    \mathcal{H}_{k,Flo} \hat{=} \left( \begin{array}{cc}
 \mathbf{0}   &   h_{Flo,-}  \\
 h_{Flo,+}    &    \mathbf{0}
\end{array}\right) , 
\end{equation}
where
\begin{eqnarray}
    h_{Flo,\pm} &=& \frac{\omega}{2} \xi_z +(J_1+J_2 \cos(k)) \pm \mathrm{i} J_2 \sin(k) \eta_y \zeta_y +E_Z \zeta_x \nonumber \\
    && +2\Delta \left(\zeta_x \xi_x + \zeta_y \xi_y\right) .
\end{eqnarray}
We may now define the normalized winding number \cite{Bomantara23}
\begin{equation}
    w_{\rm Flo, \pm} = \frac{1}{4\pi \mathrm{i} \mathrm{Tr}(\mathcal{I}_f)} \oint  \mathrm{Tr}\left(h_{\rm Flo,\pm}^{-1}dh_{\rm Flo,\pm}\right) ,
\end{equation}
where $\mathcal{I}_f$ is the infinite dimensional identity matrix acting on the frequency space. By numerically calculating $w_{\rm Flo, \pm}$ over a large range of $\Delta$ values in Fig.~\ref{fig:classresults}(d), we verify that the time-periodic Hamiltonian is indeed topologically trivial, which is also consistent with the absence of edge states in Fig.~\ref{fig:classresults}(a,b).

\section{Discussion}
\label{discuss}

\subsection{Robustness against spatial disorder}

One attractive feature of topological edge modes is their resilience against perturbations and system imperfections. To verify that our quantum-vacuum-\RB{protected} edge polaritons also display such robustness, we consider the most common type of system imperfections, i.e., the presence of spatial disorder, which in our system could be simulated by making the values of \RB{some or all} system parameters $J_1$, $J_2$, $E_Z$, and $\Delta$ site dependent. Specifically, for each lattice site $j$, we draw the value of the parameter $P\in \left\lbrace J_1, J_2, E_Z, \Delta \right\rbrace$ from the uniform distribution $\left[ \overline{P} -\delta P \overline{P}, \overline{P} +\delta P  \overline{P} \right]$.

\begin{center}
\begin{figure}
    \includegraphics[scale=0.22]{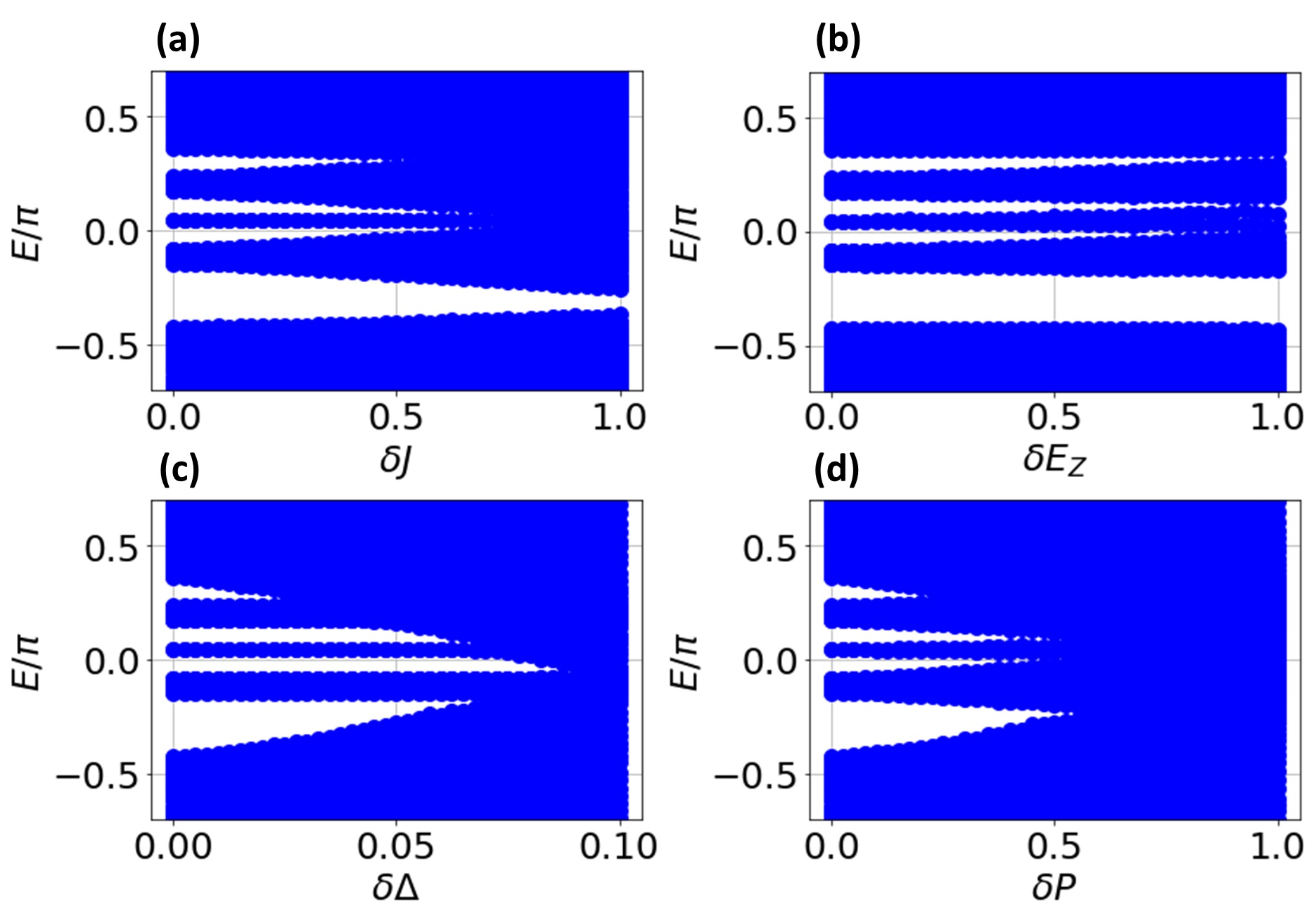}
    \caption{The disorder-averaged energy spectrum of Eq.~(\ref{SSH_vac}) for $N=16$ under OBC. Spatial disorders are considered in (a) the hopping parameters $J_1$ and $J_2$ only with $\delta J_1=\delta J_2=\delta J$, (b) the magnetic field amplitude $E_Z$ only, (c) light-matter interaction strength $\Delta$ only, and (d) all parameters with $\delta J_1=\delta J_2= \delta E_Z = 10 \delta \Delta =\delta P$. In all panels, $50$ disorder realizations are considered, $\omega=2\pi$ is taken, the Fock space is truncated to contain up to $N_p=16$ photon sectors, and the mean value of each parameter is taken as $\overline{J}_1=0.1$, $\overline{J}_2=0.5$, $\overline{\Delta}=10$, and $\overline{E}_Z=1.5$. }
    \label{fig:disorderresults}
\end{figure}    
\end{center}

Our results are summarized in Fig.~\ref{fig:disorderresults}. \RB{To understand the impact of disorder on each parameter individually, disorder is only applied on one parameter in panels (a)-(c), i.e., hopping parameters alone in panel (a), magnetic field strength alone in panel (b), and light-matter interaction alone in panel (c). For completeness, we also consider the effect of disorder on all parameters simultaneously in panel (d), which is particularly relevant in experiments where fine-tuning all parameters exactly to a single set of values is typically impossible.} As expected, the obtained quantum-vacuum-\RB{protected} edge polaritons (the in-gap flat line at almost zero energy) are evidently robust against moderate spatial disorders \RB{in all scenarios} (up to $50\%$ disorder in $J_1$, $J_2$, and/or $E_Z$ and $5\%$ disorder in $\Delta$). At large enough disorder, the band gap around zero energy closes, thus causing any edge modes to either disappear or lose their topological robustness. 

\RB{In Fig.~\ref{fig:disorderresults}(a,b), it is observed that the disorder-induced band gap closing occurs at a slightly smaller disorder strength for disorder in hopping parameters only as compared to disorder in the magnetic field. This could be attributed to the fact that the hopping parameters directly affect the topological phase of each SSH chain, whereas the magnetic field only serves as a perturbation that generally preserves the system's topology.} Finally, the smaller resistance to disorder in $\Delta$ as compared with disorder in the other parameters is attributed to the fact that $\Delta$ is one order of magnitude larger than the other parameters; consequently, $5\%$ disorder in $\Delta$ leads to terms that are of the same order of magnitude as $50\%$ disorder in the other parameters.

\begin{center}
\begin{figure}
    \includegraphics[scale=0.25]{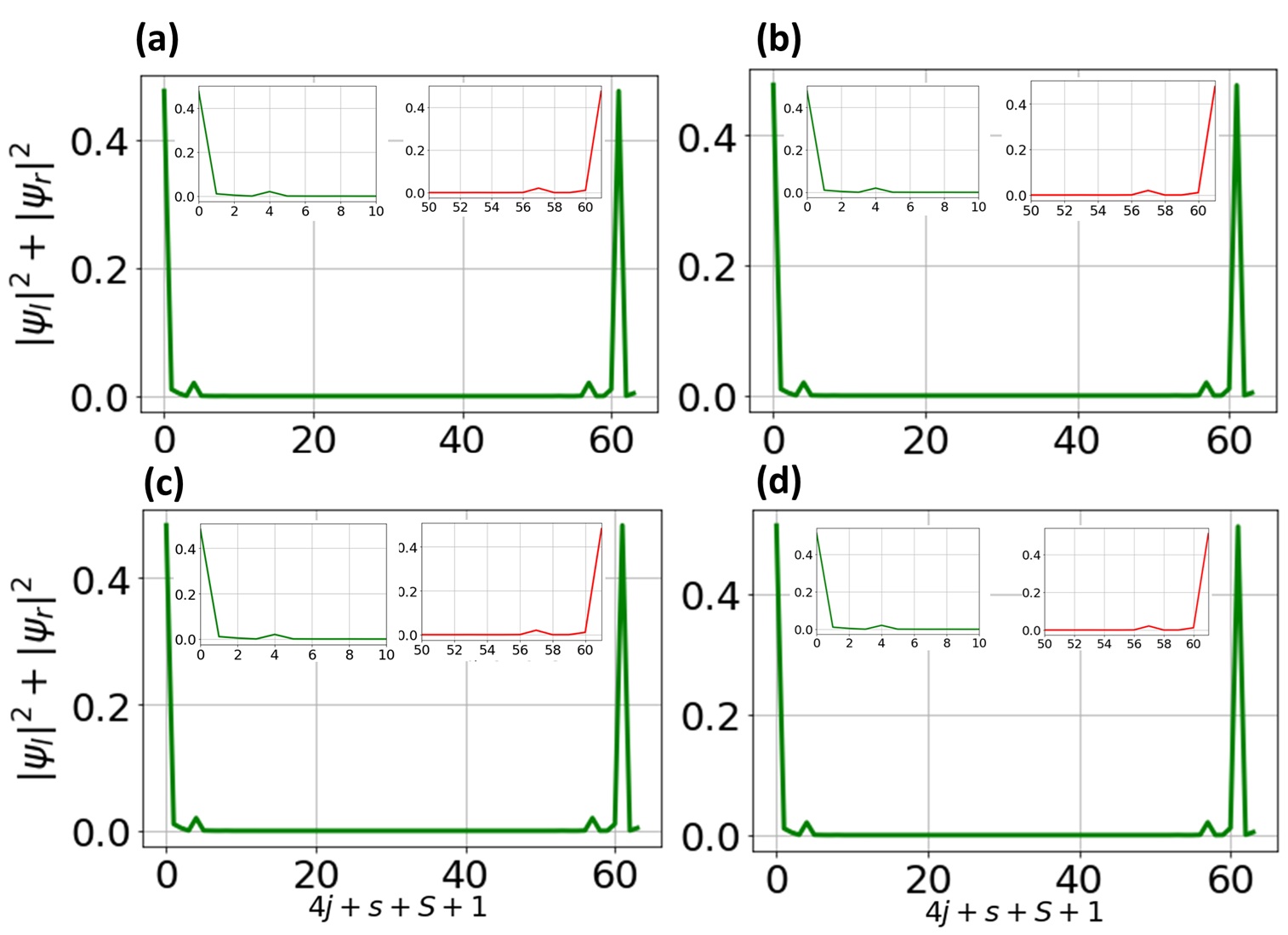}
    \caption{\RB{The disorder-averaged ``wave function" profiles \RB{(see Eq.~(\ref{wf}) and its accompanying paragraph for the precise definition)} associated with quantum-vacuum-\RB{protected} edge polaritons in the presence of disorder on the parameters considered in Fig.~\ref{fig:disorderresults}. For the cases of disorder in the hopping parameters (panels (a,d)) and/or the magnetic field (panels (b,d)), the disorder strength is set at $0.1$. Meanwhile, for the cases of disorder in the interaction strength (panels (c,d)), the disorder strength is set at $0.01$. The insets depict the zoomed-in version of each edge polariton that highlights the location of its largest peak. All other parameters are the same as in Fig.~\ref{fig:disorderresults}} }
    \label{fig:disorderresults2}
\end{figure}    
\end{center}

In Fig.~\ref{fig:disorderresults2}, we further compute and present the disorder-averaged ``wave function" profiles associated with the two quantum-vacuum-\RB{protected} edge polaritons \RB{for each disorder scenario considered in Fig.~\ref{fig:disorderresults}}. There, we have defined the ``wave function" squared as
\begin{equation}
    |\psi|^2(4j+s+S+1) = \sum_{n=0}^{N_{P}} |
 \langle s,n,S,j | \psi \rangle |^2, \label{wf}
\end{equation}
where $s=\pm 1$ is the internal level state, $S=0$ ($S=1$) for sublattice A (B), $N_p$ represents the truncation in the Fock space to enable numerical computations, and $|\psi\rangle$ is the eigenstate corresponding to the \RB{quantum-vacuum-protected} edge polariton under consideration. \RB{The combination $4j+2s+1+S$ in Eq.~(\ref{wf}) is chosen so that every $(j,s,S)$ is mapped to a unique integer, e.g., $(0,-1,0)$ is mapped to $0$, $(0,-1,1)$ is mapped to $1$, $(0,1,0)$ is mapped to $2$, etc. By closely inspecting the insets of any panel in Fig.~\ref{fig:disorderresults2}, it is observed that the edge polariton localized on the left end has the largest support on $4j+s+S+1=0$, which corresponds to the first unit cell at sublattice $A$ with internal level $-1$. Meanwhile, the edge polariton localized on the right end has the largest support on $4j+s+S+1=61$, which for $N=16$ corresponds to the last unit cell at sublattice $B$ with internal level $-1$.} Remarkably, this is consistent with the idealized schematics shown in Fig.~\ref{fig:scheme}(a) even though the system is far away from the ideal case since $J_1$ and $E_Z$ are both nonzero, and all parameters become site-dependent due to the presence of disorder. Moreover, there remains exactly two edge polaritons near zero energy, which is consistent with the topology of the system.

%Finally, it is worth noting that the emergence of small but still significant peaks near the left (right) edge for the right-edge-(left-edge-)localized polariton is simply attributed to a numerical artefact arising from the disorder averaging process. Specifically, due to the degeneracy between the two types of edge polaritons, some left-edge-localized polaritons from different disorder realizations might be included in the disorder average computation of the right-edge-localized polariton and vice versa. Nevertheless, there remains exactly two edge polaritons near zero energy as is consistent with the topology of the system.   

\subsection{Robustness against counterrotating term effect}

The TC model typically describes some physical systems under the assumption of rotating wave approximation (RWA). It amounts to approximating light-matter interaction of the form $(a+a^\dagger)\sigma_x$, which couples a particle's dipole moment with the electric field, by a theoretically simpler form $a \sigma_+ +a^\dagger \sigma_-$, where $\sigma_\pm = \sigma_x \pm \mathrm{i} \sigma_y$ \cite{Tavis68}. While RWA often serves as a good approximation to some physical systems, there are also instances in which RWA yields ``incorrect" phenomena which are otherwise absent when the exact model is considered \cite{Wang73,Rzazewski75,Larson12}.

\begin{center}
    \begin{figure}
        \includegraphics[scale=0.23]{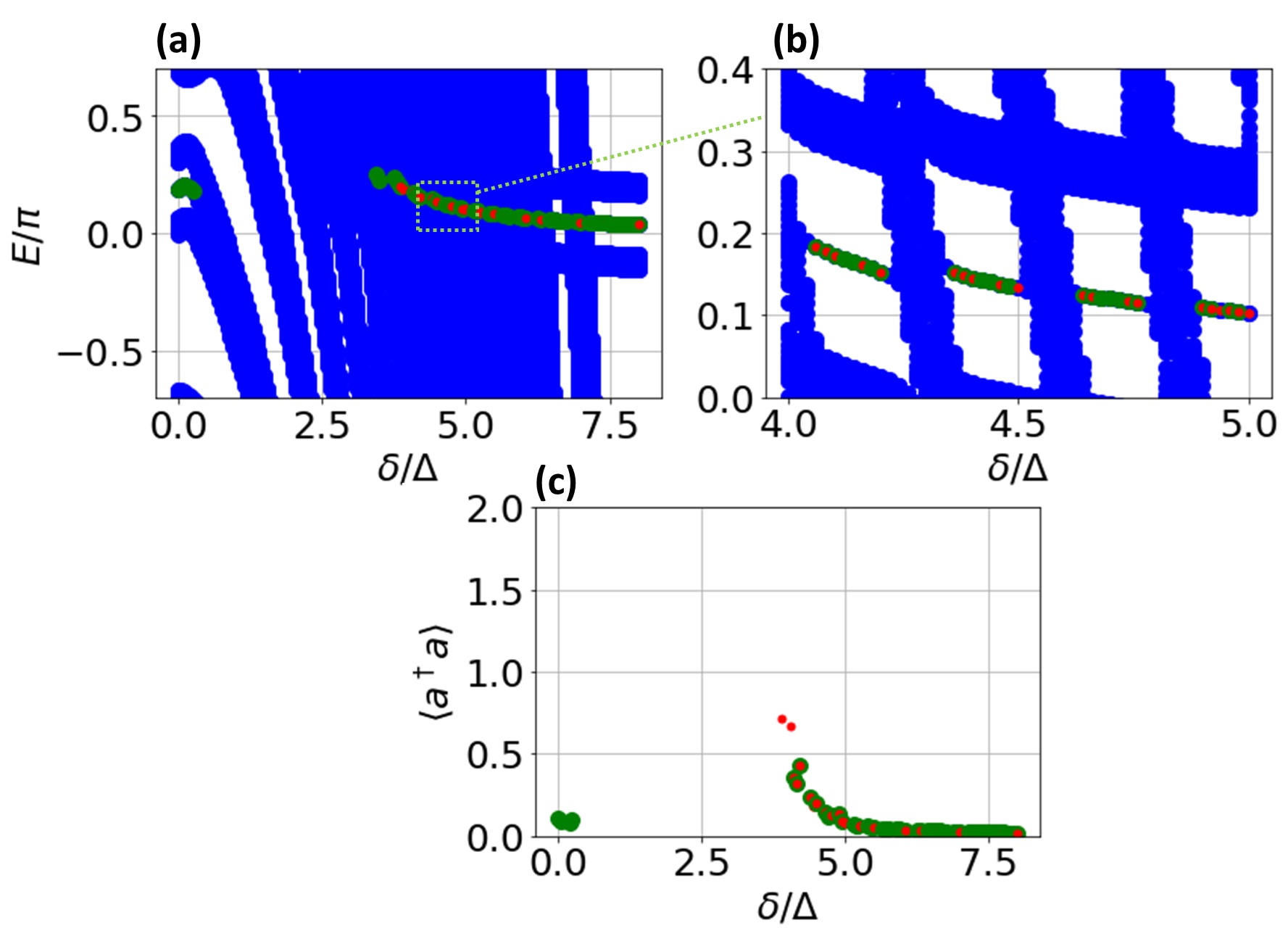}
        \caption{\RB{(a) The energy spectrum of the system for $N=20$ under OBC at varying $\delta/\Delta$ while keeping $\Delta=4$ fixed. (b) A zoomed-in view of the energy spectrum between $\delta/\Delta=4$ to $\delta/\Delta=5$. The overlapping green and red marks highlight the quantum-vacuum-protected edge polaritons localized near the system's left and right ends respectively. (c) The mean photon number for the obtained quantum-vacuum-protected edge polaritons. In all panels, the system parameters are taken as $J_1=0.1$, $J_2=0.5$, $E_Z=1.5$, $\omega=2\pi$, and the Fock space is truncated to contain up to $N_p=20$ photon sectors.} }
        \label{fig:counteresults2}
    \end{figure}
\end{center}

In view of the above, we investigate the effect of the ``counterrotating" interaction term on the formation of quantum-vacuum-\RB{protected} edge polaritons in our model. To this end, we add
\begin{equation}
    H_{\rm int}' = \sum_{j=1}^N \delta \left\lbrace a(c^\dagger_{-,B,j}c_{+,A,j}+c^\dagger_{-,A,j}c_{+,B,j})+h.c.\right\rbrace 
\end{equation}
to Eq.~(\ref{SSH_vac}). There, we have introduced a new parameter $\delta$ to describe the strength of the counterrotating term. \RB{In Fig.~\ref{fig:counteresults2}(a,b), we plot the system's energy spectrum at varying $\delta/\Delta$ while fixing $\Delta$. The quantum-vacuum-protected edge polaritons are identified as either left- or right-end localized edge modes (marked by green or red marks in Fig.~\ref{fig:counteresults2}(a,b) respectively) with almost zero mean photon number (as confirmed in Fig.~\ref{fig:counteresults2}(c)). There, we observe that quantum-vacuum-protected edge polaritons, exist over a significant range of parameter values with $\delta/\Delta \lessapprox 1/4$ or $\delta/\Delta \gtrapprox 4$. That is, quantum-vacuum-protected edge polaritons are observed not only when $\delta$ is sufficiently small, but also when $\delta$ is very large. This observation is attributed to the fact that the counterrotating interaction term also leads to the pairing between two atom-photon sectors, i.e., $(-,n)$ and $(+,n+1)$. Therefore, in the absence of the rotating term $(\Delta=0)$, the quantum vacuum sector $(+,0)$ is left uncoupled and has the potential for supporting edge modes. When $\delta$ and $\Delta$ are both nonzero, there is a competition between the different pairing mechanisms induced by both interaction terms. Consequently, quantum-vacuum-protected edge polaritons are expected to be present only when one interaction type (rotating or counterrotating) is sufficiently more dominant than the other. }

\RB{That the quantum-vacuum-protected edge polaritons are present over a window of parameter values rather than at a specific value of either $\delta=0$ or $\Delta=0$ is attributed to their topological nature. The latter is further supported by the fact that the disappearance of the quantum-vacuum-protected edge polaritons is accompanied by an energy gap closing, as also seen in Fig.~\ref{fig:counteresults2}(a,b). In particular, Fig.~\ref{fig:counteresults2}(b) highlights the fact that the quantum-vacuum-protected edge polaritons indeed arise inside a small but finite energy gap between $\delta/\Delta=4$ and $\delta/\Delta=5$, which is not visible from Fig.~\ref{fig:counteresults2}(a). } 

\begin{center}
\begin{figure*}
    \includegraphics[scale=0.45]{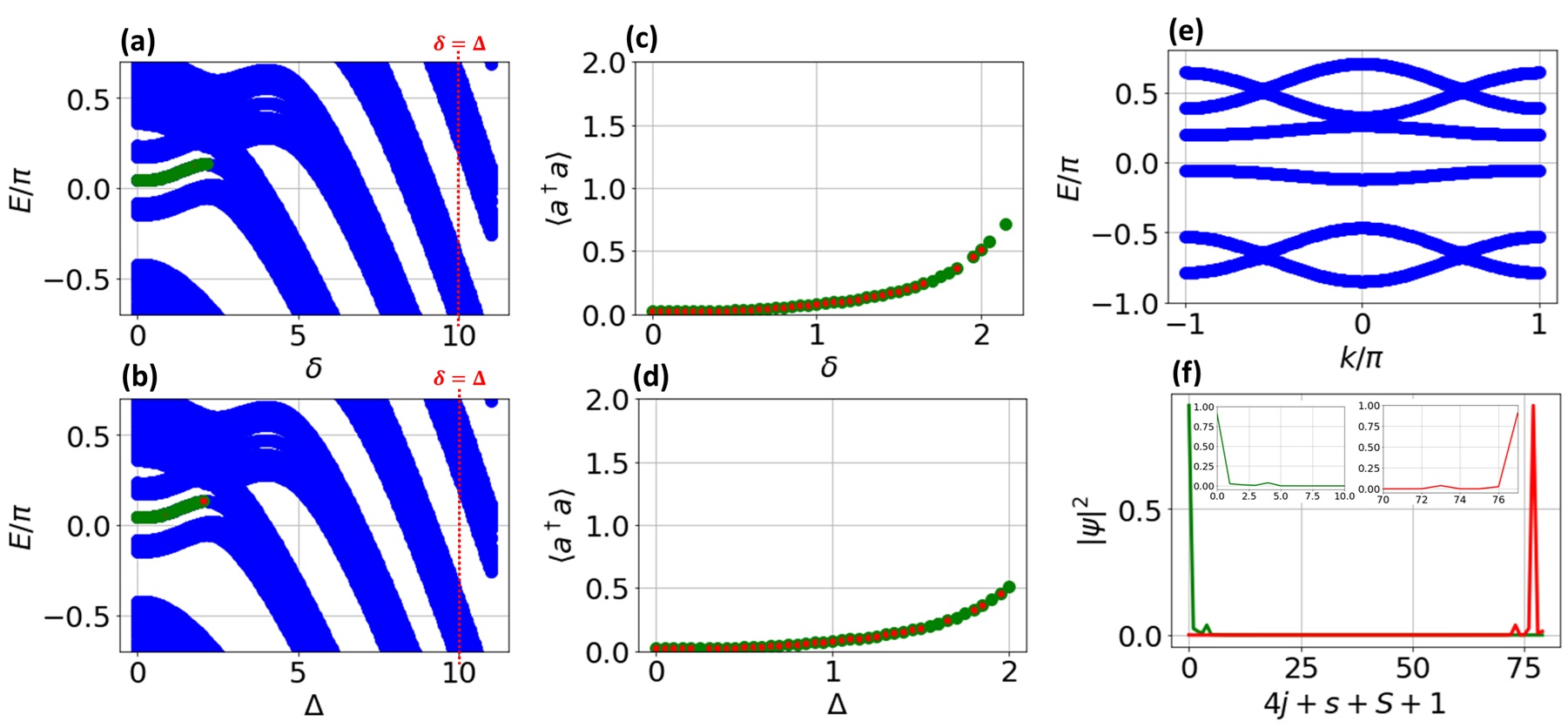}
    \caption{\RB{(a,b) The energy spectrum of Eq.~(\ref{SSH_vac}) for $N=20$ under OBC at (a) varying counterrotating interaction amplitude $\delta$ and a fixed $\Delta=10$ and (b) varying rotating interaction amplitude $\Delta$ and a fixed $\delta=10$. The overlapping green and red marks highlight the quantum-vacuum-\RB{protected} edge polaritons localized near the system's left and right ends respectively. (c,d) The mean photon number for the quantum-vacuum-\RB{protected} edge polaritons associated with (a,b) respectively.} (e) The corresponding momentum space energy bands near $E=0$ at a fixed $\delta=1$ and $\Delta =10$. (f) The ``wave function" profiles associated with quantum-vacuum-\RB{protected} edge polaritons at a fixed $\delta=1$ and $\Delta =10$ (the insets depict the zoomed-in version of each edge polariton that highlights the location of its largest peak). All other parameters are chosen to be the same as in Fig.~\ref{fig:counteresults2}.}
    \label{fig:counteresults}
\end{figure*}    
\end{center}

\RB{In Fig.~\ref{fig:counteresults}(a,b), we further plot the system's energy spectrum at varying $\delta$ and a fixed $\Delta$, and vice versa. As expected, the spectra look virtually identical and the quantum-vacuum-protected edge polaritons are observed for $\delta/\Delta \lessapprox 1/4$ or $\delta/\Delta \gtrapprox 4$ (topologically nontrivial regime). In Fig.~\ref{fig:counteresults}(c,d), we verify that the mean photon number corresponding to each quantum-vacuum-protected edge polariton is indeed very close to zero deep in the topologically nontrivial regime, consistent with the fact that such modes are close to being in a vacuum photon state. Near the topological phase transition, the mean photon numbers of the quantum-vacuum-\RB{protected} edge polaritons increase exponentially before they disappear completely. This is analogous to the behavior of an ordinary topological edge state that becomes more delocalized in a physical space as it approaches a topological phase transition point.} 

%As demonstrated in Fig.~\ref{fig:counteresults}(a), the above obtained quantum-vacuum-\RB{protected} edge polaritons persist over a significant range of $\delta$ values. In Fig.~\ref{fig:counteresults}(b), \RB{we further plot the energy spectrum and the corresponding quantum-vacuum-\RB{protected} edge polaritons at varying $\Delta$ while fixing $\delta$. That Fig.~\ref{fig:counteresults}(a) and Fig.~\ref{fig:counteresults}(b) yield virtually identical results is expected since the counterrotating interaction term also leads to the pairing between two spin-photon sectors, i.e., $(-,n)$ and $(+,n+1)$. In the absence of the rotating term, the other quantum vacuum sector $(+,0)$ is left uncoupled and has the potential for supporting edge modes.} At $\delta/\Delta \sim 1/4$, the two center bands corresponding to the zero photon sector closes, thus signifying a topological phase transition to a topologically trivial regime that does not support quantum-vacuum-\RB{protected} edge polaritons. 

%we find that the mean photon number associated with the quantum-vacuum-\RB{protected} edge polaritons found in Fig.~\ref{fig:counteresults}(a) is very close to zero at small to moderate values of $\delta$, consistent with the fact that they are close to being in a vacuum photon state. Near the topological phase transition, the mean photon number of the quantum-vacuum-\RB{protected} edge polaritons increases exponentially with $\delta$ before they disappear completely. 

Under PBC, the momentum space energy bands could be computed to uncover additional properties of the system in the presence of $\delta$. In particular, Fig.~\ref{fig:counteresults}(c) reveals a significant energy gap between the two middle bands, which are associated with the lowest (almost zero) mean photon number. Such an energy gap enables the topological protection of the resulting edge states, i.e., the quantum-vacuum-\RB{protected} edge polaritons, in the corresponding open chain setting. By contrast, the other pairs of energies shown in Fig.~\ref{fig:counteresults}(c), which correspond to larger mean photon numbers, are gapless. This in turn prevents the existence of additional topologically protected edge states in the system when OBC are applied. 

For completeness, in Fig.~\ref{fig:counteresults}(d), we present the ``wave function" profiles associated with the two quantum-vacuum-\RB{protected} edge polaritons using Eq.~(\ref{wf}). Note that both profiles look qualitatively similar to those of Fig.~\ref{fig:disorderresults2} despite the incorporation of a moderate counterrotating interaction parameter $\delta$ instead of spatial disorder. This result further highlights the topological nature of our quantum-vacuum-\RB{protected} edge polaritons.   

\RB{We close this section by highlighting the fact that since our system is in the topologically trivial regime at $\delta = \Delta$, our system cannot simply be understood as an RWA description of some physical system that only contains the usual dipole-electric-field type of light-matter interaction (corresponding to the ideal $\delta=0$). In this case, an additional type of light-matter interaction term should be added to the system to generate a difference between the rotating and counterrotating interaction amplitudes without imposing RWA. While obtaining such an additional interaction term is a nontrivial task, it is not impossible given the rapid progress in circuit quantum electrodynamics and artificial atoms technology \cite{Fink09}. Finally, it is also worth mentioning that due to the topological nature of our system, any fine tuning in generating the required interaction term is unnecessary.}

%\subsection{Quantum state transfer application (can keep J1 and J2 constant?)}

\section{Concluding remarks}
\label{conc}

In this paper, we uncovered the role of quantum vacuum in generating a pair of topologically protected edge states, termed quantum-vacuum-\RB{protected} topological edge polaritons, when a single \RBn{two-level neutral atom or artifical atom} living on a 1D lattice is subject to a TC-like coupling with a single photonic mode. The intuition behind our model construction was first elucidated by considering special parameter values and analytically identifying the exact quantum-vacuum-\RB{protected} topological edge polaritons. We then demonstrated the robustness of the obtained edge polaritons by performing numerical studies on more general parameter values, as well as by subjecting the system to spatial disorder and counterrotating interaction effect. We further highlighted their quantum vacuum nature by demonstrating that such edge states yield almost zero mean photon number and are otherwise absent in the limit of classical light. 

In summary, the quantum-vacuum-\RB{protected} topological edge polaritons discovered in this paper are made possible due to the fact that the Fock space is bounded from below by the zero photon state, which in turn mimics a physical boundary's ability of capturing topological effects. It is expected that this simple insight could be adapted to construct a variety of novel topological phases that incorporate Fock space as an artificial dimension. For example, it is imagined that a suitable light-matter interaction could be devised to construct the 1D chain analogues of the Chern insulator which supports a unidirectional-moving eigenstate from one edge to the other at very low photon mean number. The possibility and detailed analysis of such systems deserve a separate study that is best left for future work. 

\RB{Even in the context of the quantum-vacuum-protected edge polaritons discovered here, it is also interesting to explore if they could arise under a different type of light-matter interaction. In particular, the TC-like interaction considered in this paper is typically only relevant for neutral particles, e.g., atoms and excitons. For charged fermions, a more appropriate light-matter interaction is that which arises via Peierls substitution, e.g., in the spirit of Refs.~\cite{Schlawin22,Dmytruk22,Bacciconi24}. Such an interaction is not expected to yield quantum-vacuum-protected edge polaritons when directly applied to the particle Hamiltonian considered here, since it only leads to terms that couple either (pseudo)spin and photonic degrees of freedom or sublattice and photonic degrees of freedom as opposed to the interaction term considered in this paper that couples all pseudospin, sublattice, and photonic degrees of freedom simultaneously. Nevertheless, applying Peierls substitution on a different fermionic lattice model could potentially generate the necessary fermion-photon interaction that supports quantum-vacuum-protected edge polaritons and may in fact serve as a good alternative to simulate our system without RWA. To this end, the analogy with the second-order topological insulator as pointed out in this paper could be a key idea towards identifying the right fermionic lattice model, which would then constitute a promising follow-up study.}

Other potential future directions that could benefit from this paper's findings include considering the effects of non-Hermiticity, few- to many-body particle interactions, and/or time-periodicity (additional classical light). In particular, non-Hermitian topological systems form an emerging active research area in recent years that has yielded various interesting phenomena such as the so-called non-Hermitian skin effect \cite{Lee16,Alvarez18,Yao18,Kunst18,Yao18b, Li20} and topological effects on the complex energy structure \cite{Gong18, Lee19, Borgnia20,Okuma20,Zhang20}. As non-Hermiticity serves as an alternative description of open quantum systems, it could be particularly relevant in experiments involving particles in a relatively low-Q cavity. Meanwhile, the motivation for considering few- to many-body particle interaction is twofold. First, many-body interaction effect is known to enrich topological phases, which leads to the emergence of fractional quantum Hall insulators \cite{Levin09, Neupert11, Leonard23, Lu24}, Fibonacci anyons \cite{Nayak08, Lahtinen17, Trebst08, Mong14}, and parafermions \cite{Fendley12, Alicea16, Clarke13, Iemini17, Bomantara21}, among others. Second, simultaneous interaction between a single photon mode and multiple particles is known to yield superradiance, which may have nontrivial effect on the system's overall topology. Finally, time-periodicity is the main ingredient behind the so-called Floquet topological phases and their unique edge state features that have no static counterparts \cite{Bomantara16,Rudner13,Nathan15,Asboth14,Ho14,Bomantara20,Zhou18,Zhu22,Bomantara22,Koor22}. Such edge states are further 
known to be potentially advantageous for quantum computing applications \cite{Bomantara18,Bomantara18b,Bauer19,Bomantara20b,Bomantara20c,Matthies22}. Given the richness of topological phenomena induced by these effects on their own, combining them with the approach developed here is expected to be even more fruitful for uncovering the full interplay between topology and quantum vacuum.

\begin{acknowledgements}
This work was supported by the Deanship of Research
Oversight and Coordination (DROC) at King Fahd University of Petroleum \& Minerals (KFUPM) through project No.~EC221010.
 \end{acknowledgements}

\appendix

\section{Quantum-vacuum-\RB{protected} edge polaritons at $E_Z\neq 0$ and $J_1,\omega =0$ up to first order in $J_2$} 
\label{app:A}

Using $H$ in Eq.~(\ref{SSH_vac}) and the ket notation introduced in Sec.~\ref{intuition} of the main text, we first note that 
\begin{eqnarray}
    H |-,n,B,j<N\rangle &=& \sqrt{n} \Delta |+,n-1,A,j\rangle +E_Z |+,n,B,j\rangle \nonumber \\
    && + J_2 |-,n,A,j+1\rangle \label{chk1} \\
    H |-,n,A,1\rangle &=& \sqrt{n} \Delta |+,n-1,B,1\rangle +E_Z |+,n,A,1\rangle . \nonumber \\ \label{chk2}
\end{eqnarray}
We then construct the superposition 
\begin{eqnarray}
   |0_L\rangle &=& \sum_{n=0}^\infty \left\lbrace f(2n+1) |-,2n+1,B,1\rangle   - f(2n) |-,2n,A,1\rangle \right\rbrace \nonumber \\
   && + |0_L^{(1)}\rangle  
\end{eqnarray}
where
\begin{equation}
    f(n) = \left(\frac{E_Z}{\Delta}\right)^n \frac{1}{\sqrt{n!}} , \label{fn}
\end{equation}
and $|0_L^{(1)}\rangle$ is a correction term that is first-order in $J_2$. With the help of Eqs.~(\ref{chk1}) and (\ref{chk2}), it can then be shown that
\begin{eqnarray}
    H |0_L\rangle &=&  \sum_{n=0}^\infty J_2 f(2n+1) |-,2n+1,A,2\rangle  +H |0_L^{(1)}\rangle \nonumber \\
\end{eqnarray}
That is, $H |0_L\rangle$ results in a term that is first-order in $J_2$.

The right-edge-localized edge mode could be constructed in the same fashion. Specifically, we first note that 
\begin{eqnarray}
    H |-,n,A,j>1\rangle &=& \sqrt{n} \Delta |+,n-1,B,j\rangle +E_Z |+,n,A,j\rangle \nonumber \\
    && + J_2 |-,n,B,j-1\rangle \label{chk3} \\
    H |-,n,B,N\rangle &=& \sqrt{n} \Delta |+,n-1,A,N\rangle +E_Z |+,n,B,N\rangle . \nonumber \\ \label{chk4}
\end{eqnarray}
We then define 
\begin{eqnarray}
   |0_R\rangle &=& \sum_{n=0}^\infty \left[f(2n) |-,2n,B,N\rangle   - f(2n+1) |-,2n+1,A,N\rangle \right] \nonumber \\
   && + |0_R^{(1)}\rangle  
\end{eqnarray}
where $f(n)$ is defined in Eq.~(\ref{fn}) and $|0_R^{(1)}\rangle$ is a correction term that is first-order in $J_2$. Using Eqs.~(\ref{chk3}) and (\ref{chk4}), we then obtain
\begin{eqnarray}
    H |0_R\rangle &=&  -\sum_{n=0}^\infty J_2 f(2n+1) |-,2n+1,B,N-1\rangle  +H |0_R^{(1)}\rangle , \nonumber \\
\end{eqnarray}
thus resulting in a term that is first-order in $J_2$. In principle, one may continue to appropriately define $|0_L^{(1)}\rangle$ and $|0_R^{(1)}\rangle$ to obtain both zero modes with higher-order terms in $J_2$. However, we choose to stop at the first-order approximation to avoid obscuring the main physics with complex expressions.

%$|0_L^{(1)}\rangle$ is to be chosen such that $H |0_L^{(1)}\rangle$ cancels out the second term. To this end, we note that 
%\begin{eqnarray}
%    H | + , n, B, j \rangle &=& \Delta \sqrt{n+1} | - , n+1, A, j \rangle +E_Z | - , n, B, j \rangle \nonumber \\
%    && + J_2 | + , n, A, j+1 \rangle , \label{chk3}
%\end{eqnarray}
%Together with Eq.~(\ref{chk1}), we may then choose
%\begin{eqnarray}
%    |0_L^{(1)}\rangle &=& \sum_{n=0}^\infty \sum_{j} f(2n+1) \left[ \left(\frac{J_2}{\Delta \sqrt{2n+1}}\right)^{2j-1} |+,2n,B,2j\rangle \right. \nonumber \\
 %   && \left. - \left(\frac{J_2}{\Delta} \right)^{2j} \left(\frac{1}{\sqrt{2n(2n+1)} }  \right)^{2j-1} |-,2n-1,B,2j+1\rangle \right] \nonumber \\
%\end{eqnarray}

\end{document}